\documentclass[%
 reprint,
 amsmath,amssymb,
 aps,
prb,
]{revtex4-2}

\usepackage{graphicx}
\usepackage{dcolumn}
\usepackage{bm}
\usepackage{placeins}

\begin{document}

\title{Fabrication of high-quality topological insulator nanodevices from bulk-insulating air-sensitive Sb-Bi$_2$Se$_3$}

\author{Linh T. Dang}
 \affiliation{Physics Institute II, Universit$\ddot{a}$t zu K$\ddot{o}$ln, K$\ddot{o}$ln 50937, Germany}

\author{Ayushi Solanki}
 \affiliation{Physics Institute II, Universit$\ddot{a}$t zu K$\ddot{o}$ln, K$\ddot{o}$ln 50937, Germany}
 
\author{Yongjian Wang}
\altaffiliation{Current address: High Magnetic Field Laboratory of Anhui Province, HFIPS, Chinese Academy of Sciences, Hefei 230031, China.}
\affiliation{Physics Institute II, Universit$\ddot{a}$t zu K$\ddot{o}$ln, K$\ddot{o}$ln 50937, Germany} 

\author{Oliver Breunig}%
\email{breunig@ph2.uni-koeln.de}
 \affiliation{Physics Institute II, Universit$\ddot{a}$t zu K$\ddot{o}$ln, K$\ddot{o}$ln 50937, Germany}

\author{Yoichi Ando}
 \email{ando@ph2.uni-koeln.de}
 \affiliation{Physics Institute II, Universit$\ddot{a}$t zu K$\ddot{o}$ln, K$\ddot{o}$ln 50937, Germany}

\date{July 1, 2026}

\begin{abstract}
High-quality topological insulator (TI) materials are essential for the realization and detection of Majorana bound states (MBSs) in TI–superconductor hybrid platforms. Widely used compensated TIs exhibit substantial disorder and charge inhomogeneity, which may be detrimental for Majorana devices. In this regard, Sb-substituted $\mathrm{Bi_2Se_3}$ (SBS) is promising, because it is non-compensated and yet achieves very low bulk carrier density. We systematically investigate the impact of  thermal processing during microfabrication on the transport properties of SBS. We developed a room-temperature fabrication protocol that preserves the low carrier density of exfoliated SBS upon fabrication of Hall bar and nanowire devices as evidenced from the observation of quantum interference oscillations in nanowires, a large gate tunability, and clear signatures of weak antilocalization (WAL).
\end{abstract}

\maketitle
\section{Introduction}
3D TIs were predicted to host topological superconductivity when their topological surface states (TSSs) are proximitized by an s-wave superconductor \cite{fu2008}. Majorana bound states are expected to emerge at topological defects such as vortex cores, or at the ends of proximitized nanowires\cite{hosur2011,cook2011,cook2012}. Similar to other Majorana platforms such as Rashba semiconductors or iron-based superconductors, the material purity in TIs is relevant for the robust formation and detection of MBSs \cite{so2013,seongjin2021,adagideli2014,kaufhold2024}.  Disorder and inhomogeneities can introduce trivial subgap states that mimic MBS signatures or disrupt coupling of MBSs to tunnel spectroscopy probes, thus complicating the identification of MBSs \cite{pan2021}. 

In chalcogenide binary TI materials, naturally occurring crystalline defects introduce bulk carriers, often masking surface transport in magnetotransport experiments. Compensation doping in compounds such as BiSbTeSe$_2$ (BSTS) suppresses the bulk conductivity by balancing acceptors and donors through tuning of Sb/Bi and Se/Te contents. This alloying technique not only compensates native defects but also allows tuning of the Dirac point into the bulk gap.

Compensated TIs, such as BSTS, thus appear as an ideal platform for TI-superconductor hybrid devices hosting MBSs. However, inevitably arising Coulomb disorder due to compensation doping and lead to the formation of charge puddles \cite{ando2013,borgwardt2016}, posing significant challenges for the realization of Majorana states. These complications motivate revisiting binary TI materials, such as Bi$_2$Se$_3$, Bi$_2$Te$_3$ and Sb$_2$Te$_3$, among which Bi$_2$Se$_3$ is the most promising material due to its Dirac point located inside a large bulk gap \cite{zhang2009}, a key prerequisite for enhancing surface transport contributions and avoiding hybridization effects with bulk carriers.

Nevertheless, $\mathrm{Bi_2Se_3}$ is also prone to the formation of native defects \cite{navratil2004,hor2009}. As a consequence the Fermi level is frequently found close to the bottom of bulk conduction band, as observed by angle-resolved photoemission spectroscopy (ARPES) \cite{analystis2010-1,sobota2010}. Suppressing Se vacancies then becomes essential to push the Fermi level into the bulk gap, turning Bi$_2$Se$_3$ into a suitable TI material for Majorana hybrid devices. Incorporation of Sb or Ca in Bi$_2$Se$_3$ has been shown as an efficient way to reduce its carrier density  \cite{hor2009,analytis2010,checkelsky2011}. Importantly, Sb substitution is isovalent, i.e. it does not introduce compensating hole carriers, but instead alters the defect formation energy. In this regard, Sb-doped Bi$_2$Se$_3$ (SBS) appears as a prime TI-material for Majorana devices.

As MBSs are expected to emerge in nanoscale systems and require the proximity of a superconductor, nanofabrication needs to be employed in addition. Retaining the quality of SBS bulk crystals after fabrication becomes crucial for the experiment. In fact, Bi$_2$Se$_3$ is known to be extremely sensitive to the surrounding environment. Its carrier concentration increases rapidly upon exposure to oxygen, carbon monoxide and water vapor in the atmosphere \cite{analytis2010, bianchi2011,kong2011}. Several studies address the mechanism of this degradation as well as optimizing the air protection of SBS \cite{kong2011,yashina2013,benia2011,golyashov2012,park2013,green2016,atuchin2011,thomas2015,edmonds2014}. Methods such as insulating oxide capping \cite{hong2012} or post-fabrication p-type doping using F4TCNQ \cite{kim2012,cho2013,cho2015} have been employed, but they are either limited in efficiency or not compatible with device fabrication. During nanodevice fabrication, the substrate is typically exposed to elevated temperature. This can significantly degrade the electronic properties of layered topological materials \cite{breunig2021}.
Here, we systematically investigate how baking during the fabrication process influences the carrier density of SBS flakes. 
We resolved this issue by developing a baking-free fabrication protocol and demonstrated the enhanced properties of SBS nanowire devices fabricated using this method.  

\section{Quality-preserving Fabrication}
\subsection{Capping Methods}
Even though the exact degradation mechanism of Bi$_2$Se$_3$-based materials is not known, minimizing air exposure is key to maintaining low carrier density SBS. Therefore, we investigated several capping methods, including insulating oxides [atomic layer deposition (ALD)-grown Al$_2$O$_3$ and hot-wire chemical vapor deposition (HWCVD)-grown SiN$_x$], hBN encapsulation and resist covering. Fabrication was based on flakes exfoliated from a SBS bulk crystal that was grown from high-purity starting materials by the modified Bridgman technique. The carrier density of manually cleaved bulk $n_{3D}=1\times10^{17}$ cm$^{-3}$ and the mobility $\mu=9770\,$cm$^2$/Vs obtained at 2~K  compare well to previous reports on Sb-substituted Bi$_2$Se$_3$ (see Supplementary~I), indicating the high purity and low defect density of the bulk crystals \cite{analystis2010-1}. For the transport characterization of the bulk sample, we kept the time for manual cleaving and preparation of contacts prior to the measurement mentioned above as short as possible, taking about 15 minutes from glovebox storage to insertion into the cryostat's vacuum chamber. As such, we stay below the typical time scale for oxidation and band bending near the surface of Bi$_2$Se$_3$ of hours to days \cite{green2016, bianchi2011, benia2011}. Hall-bar devices were fabricated using each capping method, starting with exfoliation of SBS flakes from a bulk crystal inside an Ar glovebox, followed by immediate capping. Next, Pt/Au finger contacts were patterned using a standard electron-beam lithography (EBL) process. In this process, polymethyl methacrylate (PMMA) was spin-coated on the chip and baked at 120$^{\circ}$C for 10 min, followed by EBL exposure, development, metal sputtering, and lift-off. The resulting two-dimensional carrier densities and mobilities extracted from transport measurements (see Supplementary~V for individual Hall curves) are summarized in Table I. 
\begin{table}[h]
\caption{\label{tab:table1}
2D carrier density and mobility of SBS flakes fabricated with different capping and fabrication methods measured at 400 mK}
\begin{ruledtabular}
\begin{tabular}{lcr}
\textrm{Capping methods}&
\textrm{n$_{2\text{D}}$(cm$^{-2}$)}&
\textrm{$\mu$(cm$^{2}$/Vs)}\\
\colrule
None & 5.46$\times10^{13}$ & 1460\\
ALD & 5.84$\times10^{13}$ & 1866\\
SiN$_x$ & 3.94$\times10^{13}$ & 1951\\
hBN encapsulation& 1.96$\times10^{13}$ & 2150\\
PMMA& 2.20$\times10^{13}$ & 1880\\
PMMA without baking & & \\
\quad $V_\mathrm{G}=0$ & 1.06$\times10^{13}$& 3397\\
\quad $V_\mathrm{G}=-100$\,V & 3.76$\times 10^{12}$& 2196\\

\end{tabular}
\end{ruledtabular}
\end{table}
All capping methods slightly improved the device quality in terms of mobility compared to uncovered flakes. 
Following Ref.~\cite{kim2012} we estimate the target sheet density $n_\mathrm{2D}^*$ expected for an ideal non-degraded flake of about $10-30\,$nm thickness from our bulk crystal carrier density $n_\mathrm{3D}$. We obtain $n_\mathrm{2D}^*\sim10^{11}$--$10^{12}$\,cm$^{-2}$ as the target value to compare the different capping methods to.

While ALD and silicon nitride capping require a few seconds of air exposure for substrate loading and consequently do not help to reduce the density significantly, hBN encapsulation - often used to protect highly air-sensitive materials \cite{mayorov2011} - performs better, but the carrier density cannot be reduced down to $n_\mathrm{2D}^*$.

As an alternative to time-consuming hBN encapsulation, we also tested conventional electron beam resist (PMMA) as a capping material. It was previously suggested for protecting Bi$_2$Se$_3$ from air exposure \cite{salehi2015}. Here, in addition, we apply PMMA within a suitable glovebox in-situ after exfoliation and then spin-coat outside (see Fig.~1).  This process actually yields results similar to hBN encapsulation and, thus, was chosen as the preferred basis for a quality-preserving fabrication process.

The fact that despite capping the mobilities remain roughly within the same order and that the densities do not reach the target $n_{2D}^*$ expected from the bulk value, suggests that air exposure alone cannot account for the degradation observed during device fabrication. Among other fabrication-related factors that impact material quality, high temperature baking is known to be detrimental for certain materials, in particular to other TI compounds \cite{breunig2021}. This knowledge motivates us to investigate the effect of baking temperature on the quality of SBS flakes. From a full series of fabrication runs with different baking temperatures, ranging from 120$^\circ$C down to room temperature (i.e. no baking of the resist), we found that baking should be avoided altogether (see Appendix~B). In the following, we show that without baking and via the application of a gate voltage, we succeeded in reducing the sheet density close to the desired value $n_{2D}^*$.

\subsection{Room-temperature Fabrication Protocol}
To establish a room-temperature fabrication protocol, we replaced the baking step with vacuum-curing to harden the resist. In the absence of elevated temperature, solvent evaporation occurs more slowly, allowing the resist to gradually solidify. Because the polymer chain length and density still remain different from those of baked resist, the required dose during EBL and the  development time must be adjusted. This approach has previously been employed for the fabrication of nanodevices sensitive to high processing temperatures, such as InSb nanowire-based Majorana devices \cite{heedt2021}. In that system, elevated temperatures can induce alloying or intermixing between InSb and the proximitizing Al superconductor.

\begin{figure*}[t]
    \centering
    \includegraphics[width=\textwidth]{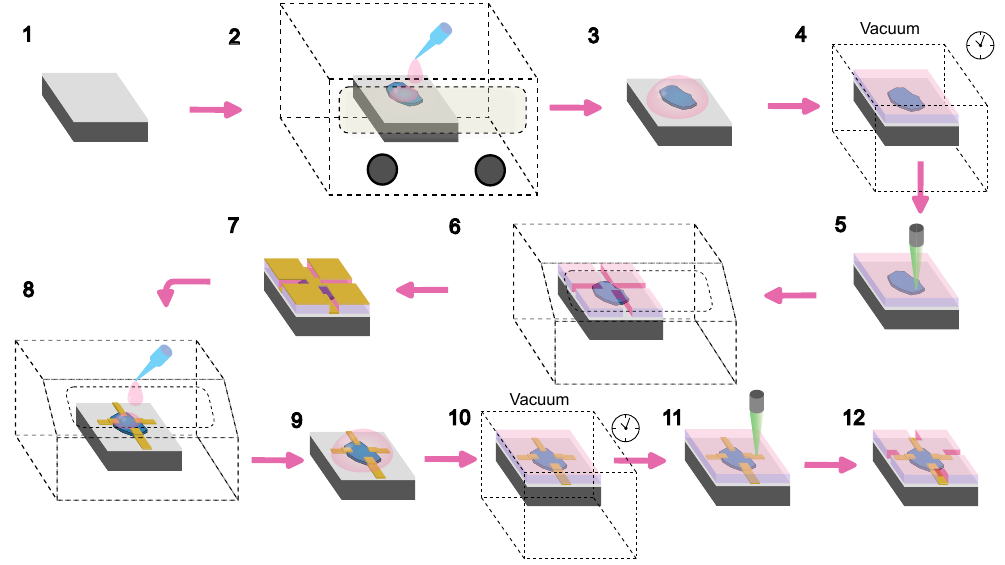}
    \caption{Room-temperature fabrication protocol for SBS nanodevices. After cleaning of the Si/SiO$_2$ substrate (1), it is brought to a glovebox (dashed lines in 2) and SBS flakes (blue) are exfoliated, followed by immediate application of a PMMA droplet which is then spin-coated under ambient conditions and vacuum-cured (3 and 4) instead of traditional high-temperature baking. After electron-beam lithography (5)  the development is performed in an N$_2$ flowbox (6) and the substrate is swiftly transferred to the vacuum chamber for metal deposition. The liftoff process and subsequent re-application of PMMA is again performed in a flowbox (8). Bonding pads are then opened following the same routine (9-12).}
    \label{fig:fig1}
\end{figure*}

A p-type Si/SiO$_2$ substrate (290 nm oxide) was cleaned with acetone, isopropanol, and O2 plasma before being transferred into the glovebox. Exfoliation of SBS flakes, followed by resist covering of the chip was carried out inside the glovebox. The chip was then removed from the glovebox, spin-coated, and placed in a vacuum chamber overnight to allow the resist to dry. Once the resist hardened, contacts were patterned by EBL. Development was performed in an N$_2$ flowbox, followed by rapid transfer to a magnetron sputtering system (Minilab, Moorfield) for metallization. Lift-off was again carried out in the N$_2$ flowbox, after which another droplet of resist was applied to protect the device. Finally, spin-coating, vacuum curing, EBL, and development were performed again to reveal pads for wire bonding (cf. Fig.~1 for the process sequence). We also optimized the EBL dose and the development time for different baking temperatures (see Supplementary~V). Upon characterizing Hall bars fabricated with different baking temperatures, we found a reduction of mobility and an increase of electron density as baking temperature increases.
\section{Transport properties of non-baked SBS flake devices}
\begin{figure*}[t]
    \centering
    \includegraphics[width=0.9\textwidth]{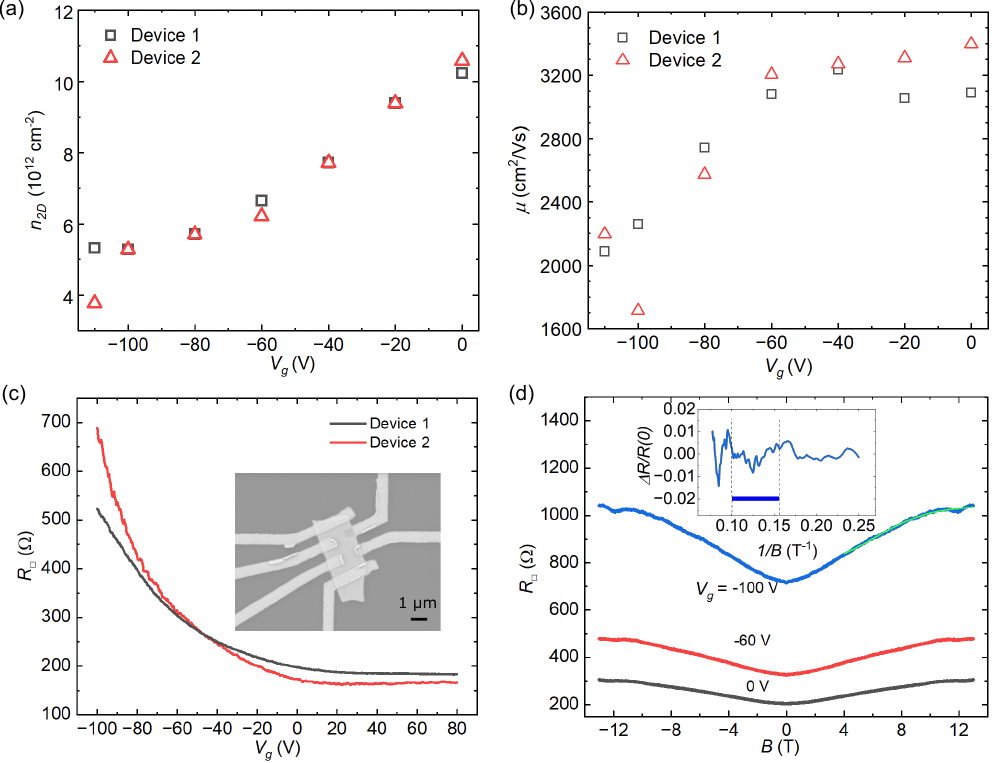}
    \caption{Transport properties of non-baked SBS Hall-bar devices. (a,b) Gate-voltage dependence of the two-dimensional carrier density and mobility for devices 1 and 2. For individual Hall curves, see Appendix~A. (c) Longitudinal resistance as a function of back-gate voltage. Inset: SEM image of a Hall-bar device fabricated without baking. (d) Magnetoresistance of device 1 measured at different gate voltages, showing the onset of oscillations at high magnetic fields. Inset: oscillation features as a function of 1/B for a gate voltage of -100 V, extracted from the high-field data (4-13 T) by subtraction of a polynomial background (thin green line in the main panel) and smoothing over 150 points. The horizontal bar denotes the oscillation period expected from the Hall density.}
    \label{fig:fig2}
\end{figure*}

Using the very same crystal used for evaluating capping methods, Hall bar devices were fabricated using the room temperature fabrication protocol described above. Transport measurements of devices were performed in a $^{3}$He cryostat (Heliox, Oxford Instruments) at 400 mK using an out-of-plane magnetic field and conventional lock-in technique. Backgate voltages were applied by electrically contacting the conductive Si substrate. The two-dimensional carrier densities of two representative devices as a function of gate voltage are presented in Fig.~2, together with resistance–gate-voltage dependence. At zero gate voltage, both devices exhibit $n_\mathrm{2D}\approx1\times10^{13}$ cm$^{-2}$, comparable to carrier densities expected when the chemical potential lies near the bulk conduction-band minimum \cite{kim2013}. Upon applying negative voltages, the carrier density can be reduced further to approximately 4-6 $\times10^{12}$ cm$^{-2}$. Such low carrier densities are typically achieved in exfoliated flakes only via external p-type doping using F4TCNQ \cite{kim2012,cho2015}, which cannot be controlled after fabrication as is the case here. 
In summary, our capping process has improved the electron density by a factor of 2 to 3 compared to unprotected processing. Additionally avoiding resist baking has improved it by another factor of 2. (cf.\ Table~I). 
Then, by gating we were approaching the order of the target $n_\mathrm{2D}^*$.
The strong gate tunability is also reflected in the longitudinal resistance increasing by a factor of three to four. This behavior suggests that the Fermi level can be tuned from the conduction-band edge into the bulk band gap.

Although a large gate tunability is observed, a clear resistance peak corresponding to the Dirac point is not reached within the accessible gate range. This behavior may result from residual Se-vacancy formation during fabrication that occurs despite all the precautions in our fabrication protocol. Another possible mechanism is band bending near the TI surface, which can lead to the formation of a trivial surface accumulation layer that becomes more and more pronounced in thinner flakes. Similar behavior has been reported for other Bi$_2$Se$_3$-based systems \cite{analystis2010-1}.

A similar gate dependence was also observed for another batch of devices with even slightly lower carrier density (see Supplementary~IV). The absence of a resistance peak in both batches suggests that density alone is not decisive for tunability through the Dirac point, but that defect states and impurity bands might play a role as well.

The electron mobility extracted for this device obtained from a non-baking process exceeds that observed for any other capping method (cf. Table I), underlining the relevance of temperature for SBS degradation. Nevertheless, the mobility of the pristine bulk crystal is not reached fully, most likely due to the exfoliation process itself or inevitable few-second air exposure after step 6 of the fabrication protocol (see Fig.~1). The decrease of mobilities upon lowering the densities at negative gate voltage (Fig.~2b) could be attributed to reduced screening of charged impurities. As the carrier density decreases, screening becomes weaker, leading to stronger scattering and reduced mobility - similar to two-dimensional electron gases (2DEGs) \cite{ihn2009}.

With the high mobilities in these devices, weak oscillatory features suggesting the onset of Shubnikov–de Haas (SdH) oscillations are observed at high magnetic fields in magnetoresistance measurements [Fig.2(d)]. While frequently found in high-quality bulk crystals like our Bi$_2$Se$_3$-crystals, quantum oscillations are rarely observed in fabricated TI Hall-bar devices, only with extremely high magnetic field and dilution refrigerator temperature \cite{qu2012}. Here, even at 400 mK weak signatures of SdH oscillations were observed at $V_G = -100 V$. At this gate voltage, the periodicity as a function of 1/B is in coarse agreement with the expectation from the Hall density (inset of Fig.~2~d).

Upon increasing the baking temperatures in the fabrication process, the mobility is reduced significantly (see Appendix~B), and consequently quantum oscillations were not observed for any device that underwent baking during the fabrication. Thus, exposure to high temperatures appears to be the main contributor to the degradation of SBS flakes. One possible mechanism for this effect is the evaporation of Se from the flake surface, which is exposed by cleavage in the van der Waals gaps, leaving behind a Bi-terminated surface. In addition, high-temperature baking of flakes may promote the diffusion of Se from the bulk to the surface, gradually introducing more Se vacancies (see Supplementary~III).
This mechanism is consistent with degradation observed in other works \cite{edmonds2014,hewitt2014,gauthier2024}. 

\section{TI nanowire fabricated at room temperature}
\begin{figure*}[t]
    \centering
    \includegraphics[width=\textwidth]{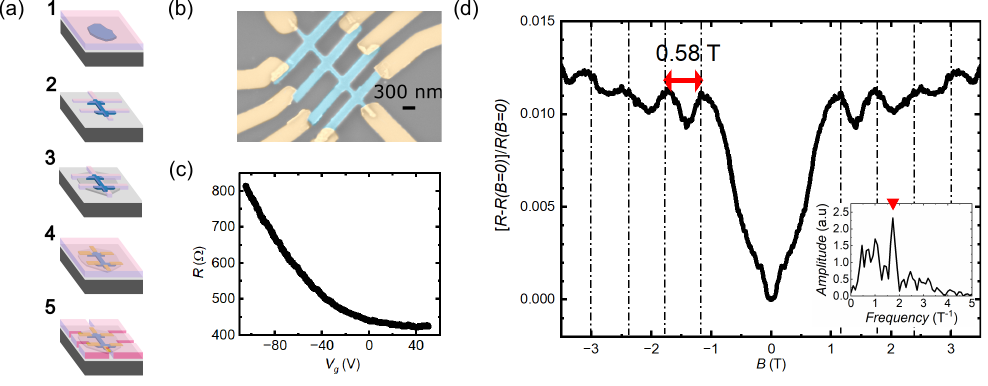}
    \caption{Nanowire device fabricated from SBS using the room-temperature protocol measured at 400 mK. (a) Simplified fabrication protocol of this nanowire device: starting with nanowire etching from flake (2), followed by SiO$_2$ deposition in the same vacuum to protect the sides of the nanowire (3). Then fabrication of the finger contacts are performed (4), finishing with opening bonding pads (5). Each fabrication uses the process described in Figure 1. (b) False-colored SEM of an etched nanowire with dimensions similar to that of the measured device. Light blue represents the etched SBS, Pt/Au contacts are shown in light yellow. (c) Gate-voltage dependence of the nanowire resistance. (d) Magnetoresistance measured at 400 mK with the magnetic field applied along the nanowire axis (symmetrized to remove a small residual Hall contribution). Dashed lines are spaced by 0.58 T, corresponding to the main oscillation also identified via Fourier transformation (Inset).  The Fourier spectrum is obtained from the full-field trace (up to $\pm13$\,T): the magnetoresistance is interpolated onto even field intervals, symmetrized, and cleared of outliers; the low-field range $|B|<2$\,T is excluded, a first-order Savitzky--Golay derivative is taken, and the result is Fourier-transformed, yielding the dominant peak at 1.73\,T$^{-1}$ (period 0.58\,T).}
    \label{fig:fig3}
\end{figure*}

Having established a fabrication process that preserves the electronic properties of bulk crystals to a large extent, an actual nanodevice was fabricated. To demonstrate the compatibility with lift-off and etching processes, we chose to fabricate a nanowire out of SBS, which is expected to show size-quantized transport and gate-tunable Aharonov-Bohm-like magnetoresistance oscillations \cite{muenning2021,kim2020,peng2009}. In addition, it forms the basis for several proposed Majorana-device architectures \cite{cook2011,cook2012,nikodem2025,nikodem2025diode}.

Each fabrication step makes use of the room-temperature protocol described above. First, the exfoliated flakes are etched into nanowire geometries of 100 to 300 nm width, with multiple contacts prepared from the flake itself used for subsequent transport measurements (called "finger contacts", similar to the geometry used in \cite{legg2022}). The etching is performed by Ar milling in the vacuum load-lock of our sputtering system (Minilab, Moorfield). Importantly, to avoid air exposure throughout the whole fabrication, immediately after etching, a 20–30 nm layer of SiO$_2$ is sputtered in-situ to cover the areas exposed by etching (Fig.~3a). In the second fabrication step, Pt/Au finger contacts are deposited to overlap the existing finger contacts. The whole chip is covered with PMMA again to prevent air exposure, with the exception of bonding pads. With this process, the main part of the TI nanowire remains protected from chemical interaction with air or Pt/Au, thereby preserving the material quality to the greatest extent possible. An SEM picture of the finished device is shown in Fig.~3b.

The longitudinal resistance of the nanowire was measured at 400 mK as a function of magnetic fields applied along the wire axis and with gate voltages applied to the substrate Si backside. Although tuning through the Dirac point has not yet been achieved, the strong gate tunability (Fig.~3c), similar to that of the Hall bars, provides first evidence that the device can be operated in the surface-transport–dominated regime. A low-field dip in the magnetoresistance is found, typically interpreted as a signature of weak antilocalization (WAL), characteristic of materials with strong spin-orbit coupling.

Beyond the WAL regime, clear oscillatory features appear at magnetic fields above 1 T. They are known to arise from phase-coherent transport in quantized Dirac subbands and have been observed in other TI nanowires, but also in SBS nanoribbons previously \cite{muenning2021,kim2020,cho2015}. We interpret them as an indication of the topological nature of the parent SBS that is preserved upon etching into nanowires. Further, the oscillations are consistent with the contribution of surface transport via Dirac subbands.

From a Fourier analysis of the full magnetic field dependence (up to $\pm 13$ T, see Supplementary~VI) we find a dominant oscillation with a frequency of 1.73 T$^{-1}$, corresponding to the most prominent oscillation with a period of 0.58 T (marked by dashed lines in Fig.~3d). Interpreting the dominant frequency as Aharonov-Bohm (AB) oscillations of period h/e yields an effective cross-sectional area $A_\mathrm{eff}=7.16\times 10^3\,\mathrm{nm}^2$, consistent with the geometric area $A_\mathrm{geom}=t\cdot w\approx7.2\times 10^3\,\mathrm{nm}^2$ with $w\simeq 120\pm 10\,$nm as obtained from SEM imaging of the fabricated device and assuming a flake thickness $t\simeq 60$\,nm which is difficult to obtain experimentally due to the employed capping method, but lies in the typical range of flake thicknesses obtained by the exfoliation method. An alternative interpretation as Altshuler-Aronov-Spivak (AAS) oscillations of period h/2e cannot be excluded. Both interpretations suggest the presence of phase-coherent transport consistent with electron trajectories encircling the nanowire perimeter.

In summary, these results indicate that the room-temperature fabrication process preserves sufficient electronic quality of SBS for mesoscopic transport and is compatible with TI nanowire–based device architectures. It maintains high electronic transport quality and does not introduce significant additional disorder or dephasing. Altogether, we conclude that high-quality nanowires can be fabricated using our fabrication protocol, enabling the fabrication of more involved hybrid TI–superconductor devices based on SBS.
\section{Conclusion}
In this work, we investigated multiple capping strategies for SBS-flake fabrication and examined the detrimental effects of high-temperature exposure during device processing. We developed a fabrication protocol that eliminates high-temperature baking and demonstrated that Sb-Bi$_2$Se$_3$ as a low-carrier-density, non-compensated TI can be a useful alternative to other TI materials which require compensation doping. In Hall-bar devices fabricated in this way the Fermi level can be tuned into the bulk band gap and we observed the onset of quantum oscillations in several SBS-flake devices. Although achieving material quality fully identical to that of bulk crystals still requires further optimization, our results demonstrate that surface-transport-dominated devices can be realized by avoiding high-temperature processing.

We further applied this protocol to fabricate a TI nanowire device, a geometry commonly used in Majorana hybrid architectures. The nanowire exhibits signatures of coherent mesoscopic transport, including weak antilocalization and quantum interference oscillations, indicating that the room-temperature fabrication protocol preserves the intrinsic electronic properties of SBS.

These results establish a viable route toward high-quality TI nanodevices based on SBS and provide a promising platform for future hybrid TI–superconductor devices aimed at Majorana experiments as well as other mesoscopic platforms utilizing Bi$_2$Se$_3$.
\begin{acknowledgments}                     
We appreciate very helpful information from Leo Kouwenhoven and Nick van Loo regarding the baking-free nanofabrications. This work was funded by the Deutsche Forschungsgemeinschaft (DFG, German Research Foundation) under Germany’s Excellence Strategy -- Cluster of Excellence Matter and Light for Quantum Computing (ML4Q) EXC 2004/2 - 390534769 and by the DFG under CRC 1238 - 277146847 (subprojects A04 and B01).

L.T.D. performed the fabrication and the transport experiments with assistance from A.S. and analyzed the data with guidance by O.B.. Y.W. grew the bulk crystals. Y.A. and O.B. conceived the concept of the work. L.T.D and O.B. wrote the manuscript with inputs of all authors.
\end{acknowledgments}

\appendix
\section{Hall data of non-baked devices}

\begin{figure}[hbtp]
  \centering
  \includegraphics[width=\columnwidth]{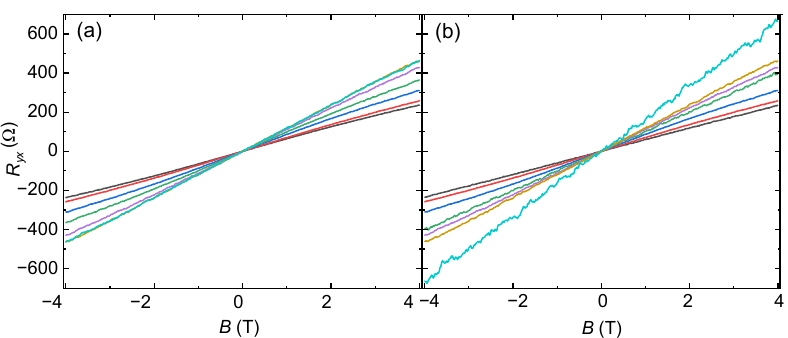}
  \caption{Hall resistance $R_{yx}(B)$ of device 1 (a) and 2 (b) as a function of out-of-plane magnetic field $B$, measured at 400\,mK for different back-gate voltages. The black, red, blue, green, magenta, dark yellow, and cyan curves correspond to $V_g = 0$, $-20$, $-40$, $-60$, $-80$, $-100$, and $-110$\,V, respectively.}
  \label{fig:fig4}
\end{figure}
Figure~4 shows the Hall resistance $R_{yx}(B)$ of devices 1 and 2 from the main text, measured at 400\,mK for a series of back-gate voltages, yielding the densities shown in Fig.~2. For both devices $R_{yx}(B)$ is linear over the full measured range of up to $\pm4$\,T, with a slope that increases monotonically with decreasing gate voltage, reflecting the reduced carrier density. The sheet carrier density and Hall mobility given in the main text are extracted by considering a single carrier type and there is no indications of another carrier contributing. We note that two channels of comparable mobility would also produce a linear $R_{yx}(B)$, so this does not by itself exclude a coexisting surface contribution. In contrast, the bulk crystal exhibits a clearly nonlinear Hall response that requires a two-carrier fit (see Supplementary~I).

\section{Carrier concentrations and mobilities of devices baked at various temperatures}
We studied the dependence of electronic properties on the baking temperature by fabrication of multiple devices in order to identify the maximum temperature that should not be exceeded during fabrication steps. Each batch of devices was fabricated using a different baking temperature for resist hardening ranging from 30 $^\circ$C up to 120 $^\circ$C, followed by overnight vacuum curing (similar to the room-temperature fabrication protocol).

For each temperature, the exposure dose and development time were adjusted to achieve the required resist-pattern resolution. The optimization procedure is described in Supplementary~V. We find a clear trend of increasing carrier density and decreasing mobility with increasing baking temperature (Fig.~5). Interestingly, the inverse dependence of the mobility is found for bulk crystals (see Supplementary~II).
We conclude that for SBS flakes, baking at conventional lithography temperatures degrades both the carrier density and the mobility. Instead, any substrate heating above room temperature should be avoided throughout the process.

Yet, the apparent finite slope of $n_\mathrm{2D}$ even at room temperature should be treated with caution due to the magnitude of statistical errors.
Even though for each temperature we fabricated up to 3 individual chips, the dependence at baking temperatures below 50 $^\circ$C may depend on further uncontrolled parameters, such that the statistical error remains rather large. Nevertheless, the difference between devices fabricated without any baking (represented by 23 $^\circ$C in Fig.~5) and those fabricated with baking is clearly visible.
As processing well below room temperature is not feasible within reasonable efforts, finally, we cannot comment on whether even lower carrier densities would be achievable via active cooling.

In contrast to flakes, for bulk crystals the effect of high-temperature exposure on the electronic properties is less pronounced (see Supplementary~II). This difference likely originates from the different surface-to-bulk ratios and the dominant role of the sample surface for Se migration and desorption. Further discussion can be found in Supplementary~III with the proposed mechanism illustrated in Fig.~S4.

\begin{figure}[htbp]
    \centering
    \includegraphics[width=0.75\columnwidth]{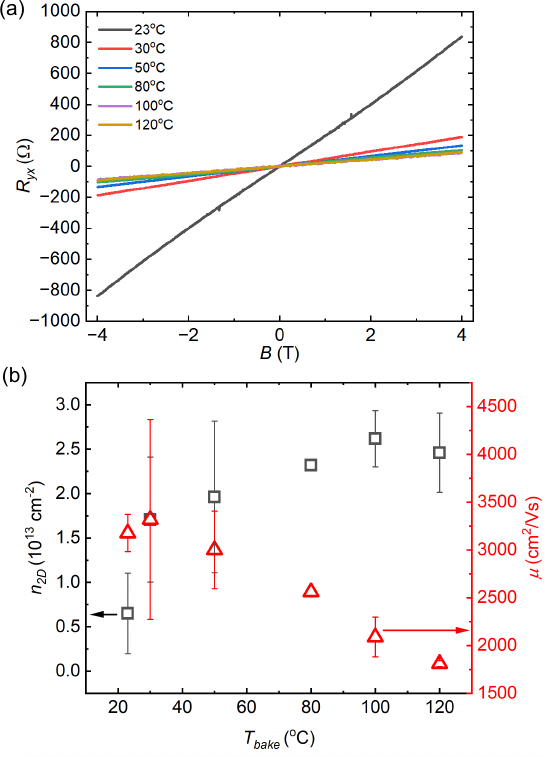}
    \caption{(a) Hall curves of exemplary devices fabricated at different baking temperatures. (b) 2D carrier concentrations (black squares) and mobilities (red triangles) of Hall bar devices fabricated with different baking temperatures. Error bars represent variation among different devices.}
    \label{fig:fig5}
\end{figure}
\FloatBarrier

\bibliography{references}

\end{document}


\title{Supplementary materials for\\ "Fabrication of high-quality topological insulator nanodevices from bulk-insulating air-sensitive Sb-Bi$_2$Se$_3$"}

\author{Linh T. Dang}
 \affiliation{Physics Institute II, Universit$\ddot{a}$t zu K$\ddot{o}$ln, K$\ddot{o}$ln 50937, Germany}

\author{Ayushi Solanki}
 \affiliation{Physics Institute II, Universit$\ddot{a}$t zu K$\ddot{o}$ln, K$\ddot{o}$ln 50937, Germany}
 
\author{Yongjian Wang\textsuperscript{*}}
\affiliation{Physics Institute II, Universit$\ddot{a}$t zu K$\ddot{o}$ln, K$\ddot{o}$ln 50937, Germany} 

\author{Oliver Breunig\textsuperscript{$\dagger$}}
 \affiliation{Physics Institute II, Universit$\ddot{a}$t zu K$\ddot{o}$ln, K$\ddot{o}$ln 50937, Germany}

\author{Yoichi Ando\textsuperscript{$\ddagger$}}
\affiliation{Physics Institute II, Universit$\ddot{a}$t zu K$\ddot{o}$ln, K$\ddot{o}$ln 50937, Germany}
\date{July 1, 2026}
\maketitle

\vspace*{\fill}

\begin{flushleft}
\footnotesize
$^{*}$ Current address: High Magnetic Field Laboratory of Anhui Province,
HFIPS, Chinese Academy of Sciences, Hefei 230031, China.\\
$^{\dagger}$ breunig@ph2.uni-koeln.de\\
$^{\ddagger}$ ando@ph2.uni-koeln.de

\end{flushleft}

\clearpage
\renewcommand{\thefigure}{S\arabic{figure}}
\renewcommand{\thetable}{S\arabic{table}}
\renewcommand{\theequation}{S\arabic{equation}}

This supplementary material provides additional characterization of the SBS bulk crystal (Supplementary I), investigates thermal annealing effects in bulk crystals (Supplementary II), discusses the temperature-induced degradation mechanism (Supplementary III), presents reproduced data in Hall-bar devices 3 and 4 (Supplementary IV) and describes optimization of the fabrication process (Supplementary V), provide full-range nanowire magnetoresistance data (Supplementary VI).

\section{bulk crystal characterization}
Our Sb-doped Bi$_2$Se$_3$ crystals were grown from a melt with a molar ratio of Sb:Bi:Se = 10:50:130 using modified Bridgman technique. This mixture was sealed in quartz tubes under a partial pressure of argon, and heated to 740 $^\circ$C and kept for 20 h with intermittent shaking to ensure homogeneity of the melt. The quartz tube was then cooled to 550 $^\circ$C in 50 h, followed by annealing for 80 h.

\begin{figure*}[b]
    \centering
    \includegraphics[width=0.8\textwidth]{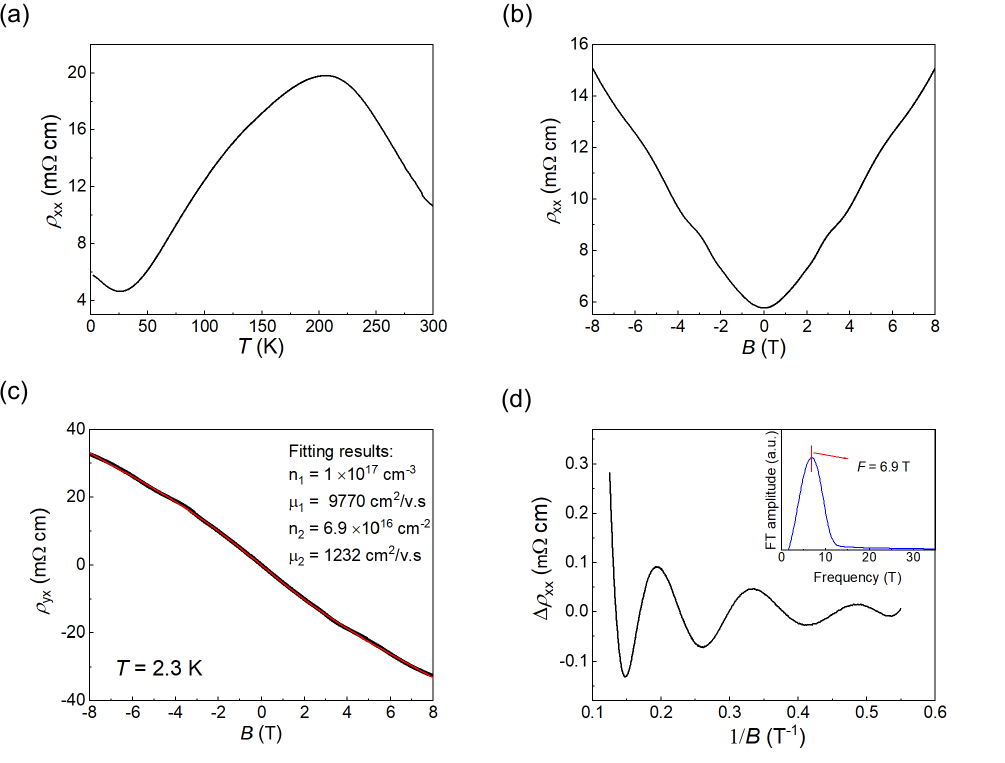}
    \caption{(a) $\rho_{xx}(T)$ curve of SBS crystal labelled 20231205-2. (b) Longitudinal resistivity as a function of out-of-plane magnetic field and (c) Hall resistivity of SBS bulk crystal flake. (d) SdH oscillations analyzed from magnetoresistance. Inset: Fourier analysis of these SdH oscillations.}
    \label{fig:figS1}
\end{figure*}

Exfoliation of the bulk crystal was carried out in an Ar glovebox to avoid air exposure, except for bulk crystal cleaving. For transport measurements, we cleaved a flake with lateral dimensions of about 100–400 $\mu$m and a thickness of 10–40 $\mu$m. The contacts allow measurements of longitudinal and Hall resistance under an out-of-plane magnetic field. The measurements were performed in a physical property measurement system (PPMS) at a base temperature of 2.1 K. Resistivity versus temperature, Hall resistance, magnetoresistance, and their analysis are shown in Fig. \ref{fig:figS1}.

The $\rho_{xx}(T)$ curve shows semiconducting behavior at high temperature, followed by a metallic regime below about 200 K and a slight upturn below 25 K. This behavior is typical for a moderately insulating topological insulator \cite{checkelsky2011,analytis2010}. From the Hall data, a two-band model fit indicates two carrier species contributing to transport, using the formula in \cite{ren2010}. The high-mobility channel has a carrier density of about 1$\times$10$^{17}$cm$^{-3}$ and mobility of 9770 cm$^2$/Vs, while the low-mobility channel has a carrier density of 6.9$\times$10$^{16}$cm$^{-3}$ and mobility of 1332 cm$^2$/Vs.

SdH oscillations are visible in the magnetoresistance. The oscillation frequency of 6.9 T corresponds to a carrier density of about 1$\times$10$^{17}$cm$^{-3}$, indicating that the high-mobility carrier species is responsible for the quantum oscillations. Overall, the carrier density and mobility confirm the high electronic quality of the starting SBS crystal, among the highest mobility values reported for Sb-doped Bi$_2$Se$_3$ crystals.
\section{High temperature annealing effect on bulk crystal}
We studied the effect of high temperature on the bulk crystal and compared it with the behavior of fabricated nanodevices. Using a protocol similar to Supplementary~I, we cleaved two flakes from the same crystal. One flake was fully covered with PMMA, while the other was left uncovered. After vacuum curing the PMMA, both flakes were loaded into the PPMS for characterization.

\begin{figure*}[h]
    \centering
    \includegraphics[width=\textwidth]{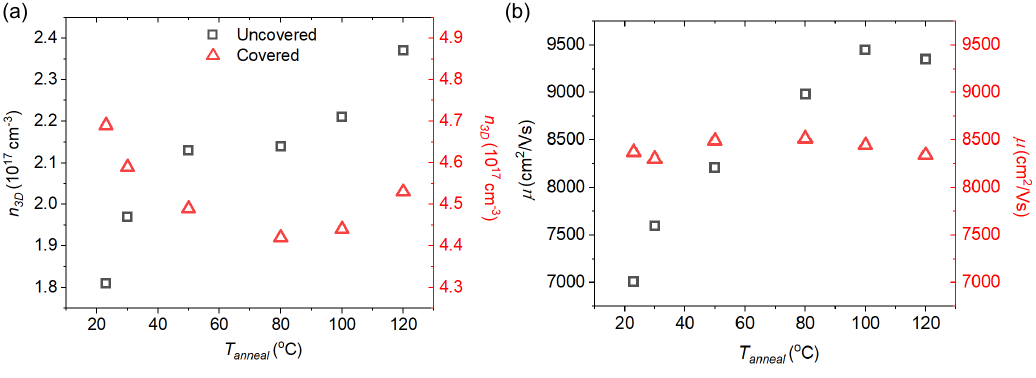}
    \caption{a) 3D carrier density and b) mobility of SBS bulk crystal as a function of annealing temperatures. The covered and uncovered datasets are plotted against independent left and right y-axes, respectively, to emphasize their temperature dependence.}
    \label{fig:figS2}
\end{figure*}

Starting from the pristine crystal, after each Hall measurement at 2.3 K the samples were annealed inside the PPMS at elevated temperature for 10 minutes and then cooled down again. With this method we can track how annealing affects the carrier density and mobility of the same crystal. The results are shown in Fig. \ref{fig:figS2}.

Similar to the fabricated devices, high temperature also increases the carrier density in the bulk crystal, although the effect is more moderate. A temperature increase of 50 K leads to an increase of about 0.5$\times$10$^{17}$ cm$^{-3}$ in carrier density. In the PMMA-covered sample, the carrier density instead decreases slightly as the annealing temperature increases. Note that the effect is cumulative: at each annealing step the sample already contains carriers introduced during the previous heating cycle. Nevertheless, it is clear that PMMA capping strongly suppresses the effect in the bulk crystal. This observation is consistent with our hypothesis of surface-level Se evaporation (discussed further in Supplementary~III).
\begin{figure*}[b]
    \centering
    \includegraphics[width=0.95\textwidth]{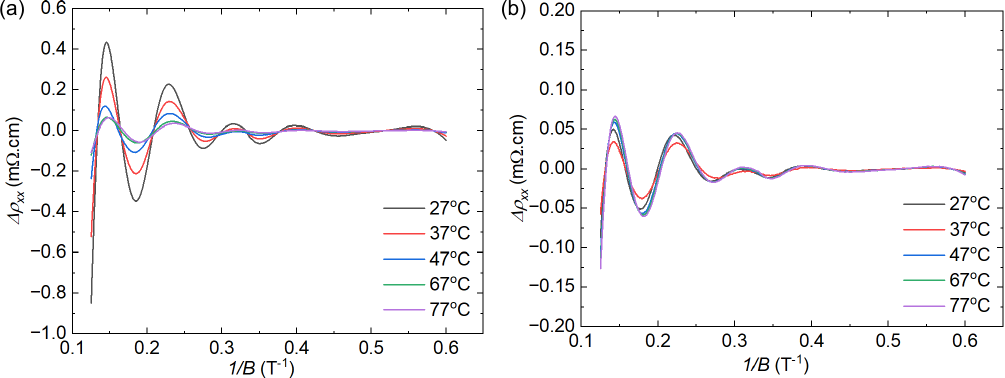}
    \caption{SdH oscillations measured after annealing at different temperatures a) without and b) with PMMA covering.}
    \label{fig:figS3}
\end{figure*}
Unexpectedly, the mobility increases upon heating in the uncovered sample, while it remains nearly unchanged in the covered sample. One possible explanation is that Se evaporation near the surface enables migration of bulk interstitial Se atoms, which reduces scattering and improves mobility. 

We also investigated quantum oscillations as a function of annealing temperature in these flakes. Unlike fabricated devices, SdH oscillations can be clearly observed in bulk crystals. Fig. \ref{fig:figS3} shows SdH oscillations measured in PMMA-covered and uncovered flakes after annealing at various temperatures. Oscillations are visible in both cases, but their amplitude decreases noticeably with increase in the annealing temperature in the uncovered sample, while changes are negligible in the covered sample.

This result appears counterintuitive, since the mobility increases with annealing. One possible explanation is that multiple transport channels contribute to conduction, but only one channel produces the high-field oscillations. Annealing may reduce the mobility of this specific channel, leading to a smaller SdH amplitude. Another possibility is that different scattering mechanisms may influence transport mobility and SdH oscillations in different ways, leading to the opposite trends observed here.

\section{Temperature-induced degradation mechanism}
\begin{figure*}[t]
    \centering
    \includegraphics[width=\textwidth]{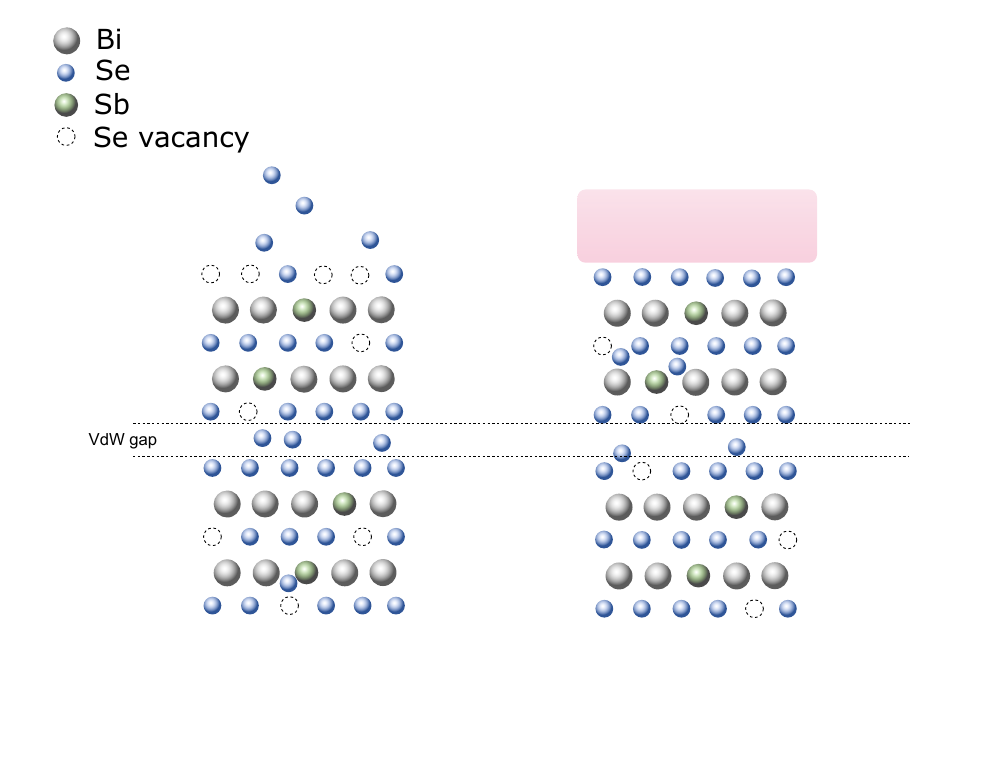}
    \caption{Evaporation of Se (light blue) atoms near the surface of SBS crystal with and without resist capping. Without capping, more Se atoms escape the crystal, leaving behind donor-like Se vacancies. There are also Se vacancies and interstitial Se, acting as scattering centers and affecting transport mobility.}
    \label{fig:figS4}
\end{figure*}

The strong difference in the mobility response to high temperature between bulk crystals (previous section) and fabricated flakes (Appendix~B, main text) remains not fully understood. Our working hypothesis is that the main degradation mechanism is the evaporation of Se near the surface of SBS. The proposed mechanism is illustrated in Fig. \ref{fig:figS4}.

This picture explains most of the behaviors observed in SBS so far. As Se atoms escape from the surface, resist capping can partially suppress the degradation. High temperature accelerates this evaporation process, leading to the formation of more Se vacancies and therefore higher carrier density. This interpretation is consistent with the superior transport properties observed in bulk crystals before annealing and in fabricated devices without baking.

In bulk crystals, although Se evaporation creates vacancies near the surface, it can also promote diffusion of bulk Se interstitial atoms toward the surface layer, partially healing the crystal and increasing the mobility. In addition, the higher carrier density enhances electron screening, which can further reduce scattering. In fabricated flakes, the situation is different. Because the surface-to-volume ratio is much larger, the vacancy-rich surface layer dominates transport. As a result, mobility decreases as the devices experience higher baking temperatures.
\begin{figure*}[ht]
    \centering
    \includegraphics[width=\textwidth]{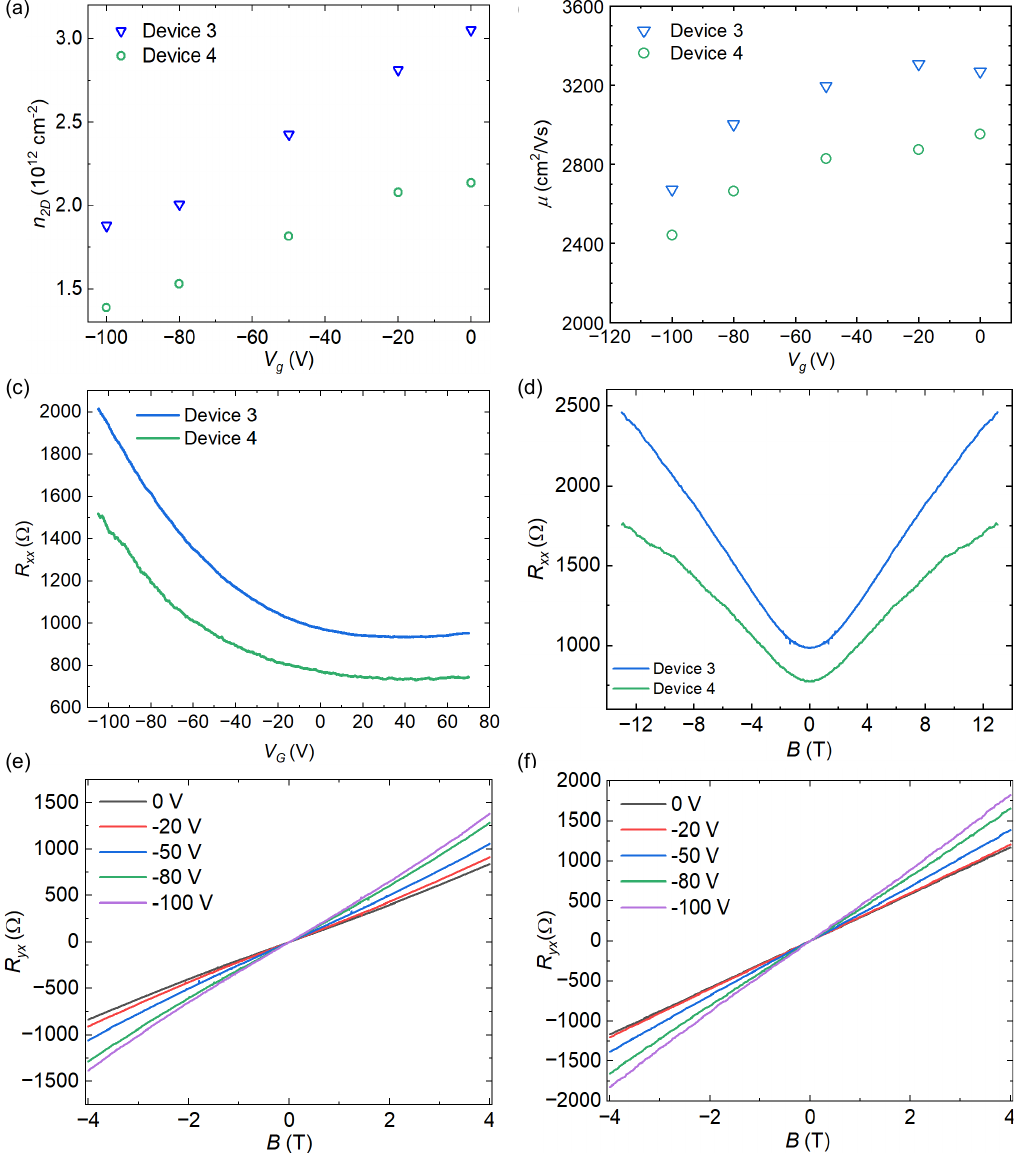}
    \caption{(a) 2D carrier density, (b) mobility and (c) longitudinal resistance of device 3 and 4 as a function of gate voltages. (d) Magnetoresistance of device 3 and 4. (e) and (f) individual Hall curves of device 3 and 4 as a function of gate voltages, respectively.}
    \label{fig:figS5}6
\end{figure*}
\section{Reproduced data in device 3 and 4}
The room-temperature fabricated devices 3 and 4 whose data are shown in Fig. \ref{fig:figS5} provide an additional confirmation of the results presented in the main text. Compared with devices 1 and 2, devices 3 and 4 exhibit even lower two-dimensional carrier density and higher mobility.

However, only a hint of quantum oscillations in terms of small wiggling of the magnetoresistance data were observed at high magnetic field ($>$10 T), as shown in Fig. \ref{fig:figS5}(d). The reason for this difference is unclear. Hall curves of devices 3 and 4 at different gate voltages are shown in Fig. \ref{fig:figS5}(e–f), respectively. From the gating curve of device 3 (Fig. \ref{fig:figS5}(c)), no Dirac point is observed, similar to devices 1 and 2. We believe this is due to the limited gating range of the experiment. With a wider gating range, the Dirac point in SBS Hall bars fabricated using the room-temperature protocol should become accessible. These results support the conclusion in the main text.
\section{Optimization of fabrication protocol}

\begin{figure*}[h]
    \centering
    \includegraphics[width=0.7\textwidth]{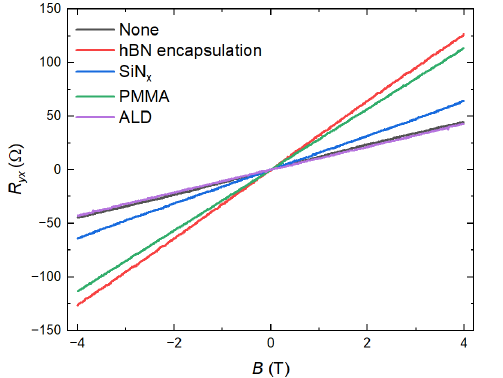}
    \caption{Hall curves of SBS-based Hall bars fabricated with different capping methods.}
    \label{fig:figS6}
\end{figure*}

Prior to the development of room-temperature protocol, different capping methods were tested to examine the protection they provided against air-exposure-induced degradation. Hall bars were fabricated from flakes with different cappings, as described in Section II of the main text. The Hall curves are presented in Fig. \ref{fig:figS6} - the extracted carrier densities and mobilities from these Hall curves are presented in Table~I of the main text.

\begin{figure*}[h]
    \centering
    \includegraphics[width=0.8\textwidth]{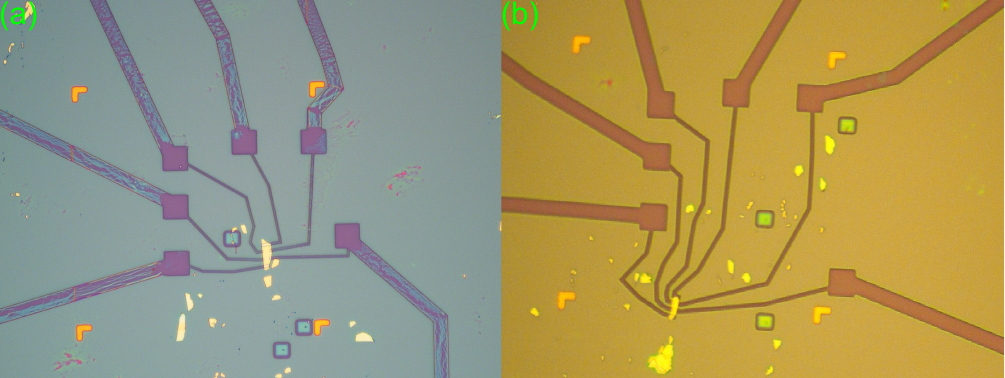}
    \caption{Exposed structures after development a) with residue (underdosed) and b) without residue (proper dose).}
    \label{fig:figS7}
\end{figure*}

As resist-capping was selected due to its simplicity and decent protection provided, optimization of the EBL exposure dose and development time for vacuum-cured PMMA is needed. Starting from an electron dose of 120 $\mu$C/cm$^{2}$ for a baking temperature of 120 $^\circ C$, the development time was adjusted for each lower baking temperature individually to optimize the development. Two different doses are used: a lower dose for inner fine structures and a higher dose for outer structures. If the exposure dose is too low, increasing the development time cannot fully remove the resist in the exposed area, leaving a permanent residue similar to that shown in Fig. \ref{fig:figS7}(a). When such residue is observed, the exposure dose is increased until the residue disappears. The development time is then optimized in the same way, resulting in a clean structure after development as shown in Fig. \ref{fig:figS7}(b).

After development, the sample is immediately transferred from the N$_2$ flowbox to the sputtering system (Moorfield Minilab S060A). Gentle Ar milling is performed in the load lock (50 W, 10 s) to remove remaining resist residue. Next, 5 nm of wetting Pt and 30–35 nm of Au are deposited. Standard lift-off in acetone with ultrasonication is then performed until the final device structure is obtained. The ultrasonication conditions are adjusted for each fabrication attempt, as the optimum also depends on the baking temperature.

After lift-off, the sample is removed from acetone, rinsed in IPA, and transferred back to the N$_2$ flowbox. The sample is then coated with PMMA again and undergoes another lithography step to open the bonding pads.

The optimization procedure described above is repeated for different baking temperatures. The resulting exposure dose and development time for each baking temperature are summarized in Table S1.

We optimized the exposure dose and the development time for both PMMA A4 950K and ZEP520A resists. Other parameters in the process remain similar to conventional bake-based protocol. Table \ref{tab:tableS1} shows the optimized dose for fine (inner) features of less than 1$\,\mu$m size and for coarse (outer) structures larger than that.
\begin{table}[h]
\caption{\label{tab:tableS1}%
E-beam exposure dose and development time for different baking temperatures
}
\begin{ruledtabular}
\begin{tabular}{cccc}
\textrm{Baking temperature (K)}&
\textrm{Inner dose ($\mu$C/cm$^{2}$)}&
\textrm{Outer dose ($\mu$C/cm$^{2}$)}&
\textrm{Development time (s)}\\
\colrule
296 & 280 & 200 & 150\\
303 & 300 & 200 & 140\\
323 & 300 & 200 & 90\\
353 & 120 & 140 & 90\\
373 & 120 & 120 & 90\\
393 & 120 & 120 & 90\\
\end{tabular}
\end{ruledtabular}
\end{table}

The room-temperature (296 K) recipe listed in Table S1 was used to fabricate the nanowire device discussed in the main text. To obtain a nanowire width of 130 nm, we used a slightly larger exposure dose (300 $\mu$C/cm$^{2}$) and set the designed width to 170 nm. After exposure, an additional overnight curing step was introduced before development to improve the solidity of the resist. This step noticeably improves the resist sidewall profile for delicate structures such as the nanowire. 

\section{Nanowire data}
\begin{figure*}[htbp]
    \centering
    \includegraphics[width=\textwidth]{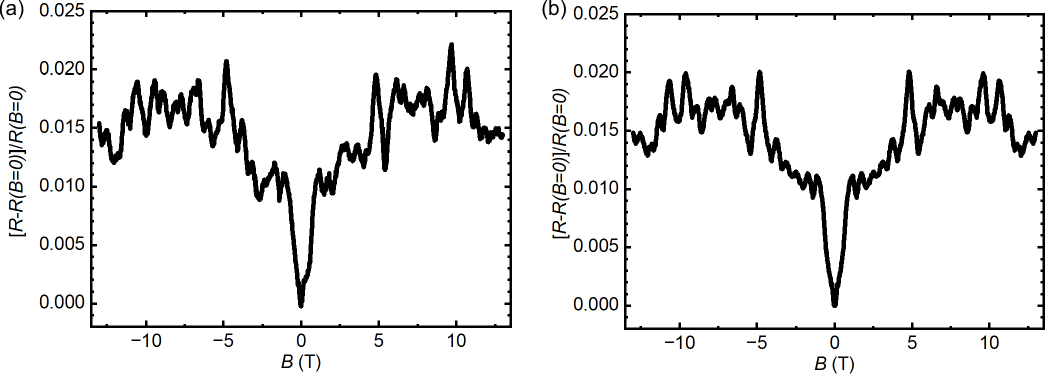}
    \caption{Longitudinal resistance of the etched nanowire fabricated without baking as function of magnetic field applied parallel to the nanowire axis, up to $\pm 13$ T: (a) raw data and (b) symmetrized data. Low-field section of panel (b) is shown in Fig. 3(d) of main text.}
    \label{fig:figS8}
\end{figure*}

Fig. \ref{fig:figS8} shows the transport data obtained over the full possible magnetic field range with a weak antilocalization dip around zero field and magnetoresistance oscillations. The data were symmetrized in order to enhance the oscillatory component. Note that this process does not introduce features on top of those already contained in the raw data, but it only removes finite asymmetric contributions introduced by an imperfect geometry of the film contacts to the nanowire. Our result are similar to the reports for VLS-grown nanoribbons from Ref.~\cite{peng2009}. Nanoribbons, however, suffer less from air-exposure degradation compared to bulk crystal. 

\bibliography{references}